\documentclass[twocolumn,amsmath,amssymb,superscriptaddress,showkeys,showpacs]{revtex4-1}
\usepackage{graphicx}
\usepackage{bm}
\usepackage{hyperref}

\begin{document}
\preprint{\today}

\title{Elongational viscosity of weakly entangled polymer melt \\ via 
coarse-grained molecular dynamics simulation}

\author{Takahiro Murashima}

\email{murasima@cmpt.phys.tohoku.ac.jp}

\affiliation{
Department of Physics, Tohoku University, \\
Sendai, Miyagi 980-8578, Japan
}

\author{Katsumi Hagita}
\affiliation{
Department of Applied Physics, National Defense Academy, \\
Yokosuka, Kanagawa 239-8686, Japan
}

\author{Toshihiro Kawakatsu}
\affiliation{
Department of Physics, Tohoku University, \\
Sendai, Miyagi 980-8578, Japan
}

\date{\today}

\begin{abstract}
We investigated the elongational flows of the weakly entangled linear polymer melt using a coarse-grained molecular dynamics simulation.
We extended the uniform extensional flow (UEF) method developed by Nicholson and Rutledge ({\it D. A. Nicholson and G. C. Rutledge, J. Chem. Phys., {\bf 145} 244903 (2016)}) for application to Langevin dynamics.
We succeeded in observing the elongational viscosity of the weakly entangled linear polymer melt from the equilibrium state to the steady state using the extended UEF method, whereas the conventional rectangular parallelepiped shape technique for extensional flows has failed to do so for over 20 years.
\end{abstract}


\maketitle

\section{Introduction}
Elongational flows are important in polymer processing~\cite{BHW1989}.
For fiber spinning, uniaxial elongational flow comprises the main part of the processing. Biaxial elongational flow appears in the film extruding process and planar elongational flow is found in a region of the cross-section during the molding process.
To enhance the performance of the products, an understanding of molecular dynamics in polymer processing is important and the molecular dynamics (MD) simulation can be helpful.
However, for over 20 years, it has been difficult to handle general elongational flows in MD simulation. 

When considering elongational flow in MD simulation,
we generally assume that the elongational axis is set parallel to the axis of the unit cell.
When we apply a uniaxial elongation to the system, for example,
the unit cell is deformed to the rectangular parallelepiped shape (RPS).
This conventional RPS method~\cite{Heyes1985} for elongational flow fails at a finite strain where the unit cell collapses and the stress diverges or rapidly damps, where the simulation breaks down.
The conventional RPS technique for the elongation is useful only for solids, not for liquids.

Using the Kraynik--Reinelt boundary conditions (KRBCs)~\cite{KR1992}, where the unit cell at the initial state is tilted in the direction of elongation,
the collapse of the unit cell in the planar elongational flow was avoided~\cite{TD1999,BC1999,HT2003}. 
In the case of planar elongational flow under the KRBCs,
the periodic copies of the origin of the unit cell correspond to the initial square unit cell at a certain strain $\epsilon_{\rm p}$, and then the collapsed unit cell can recover its original shape by switching the unit cell. This situation is similar to shear flow. 
When the shear strain $\gamma$ is equal to unity, the periodic copies of the deformed unit cell correspond to the original square lattice.
However, the KRBCs have not been considered applicable to uniaxial and biaxial elongational flows, because planar elongational flow is two-dimensional flow but uniaxial and biaxial elongational flows are three-dimensional flows.
Recently, Dobson~\cite{Dobson2014} and Hunt~\cite{Hunt2015} have generalized the KRBCs applicability to uniaxial and biaxial elongational flows and applied it to investigate the repulsive Lennard--Jones (LJ) fluid.
Nicholson and Rutledge~\cite{NR2016} developed the uniform extensional flow (UEF) algorithm on the basis of the methods developed by Dobson~\cite{Dobson2014}, and applied it to study the crystal nucleation of $n$-eicosane (${\rm C}_{20}$) melts represented by a united atom (UA) picture under shear and uniaxial elongational flows.

To investigate the entangled polymer melts, coarse-graining in time and length scales is important because of the high computational costs due to the long relaxation time of polymers.
The Kremer--Grest (KG) model~\cite{KG1990} is the representative model of coarse-grained molecular dynamics (CGMD) simulation for polymer melts,
where a polymer is represented by beads and springs, and its dynamics is described by the Langevin equation of motion.
Although we can find several works considering entangled polymer melts with UAMD
~\cite{Harmandaris2003,Harmandaris2007,Harmandaris2009,Takahashi2017,Takahashi2017b},
the CGMD is important for investigating polymer dynamics while maintaining a reasonable computational cost.
As an example, Harmandaris et al.~\cite{Harmandaris2007} reported that the time scale accessible in CGMD with the 2:1 CG model (one monomer is mapped onto two CG particles) of polystyrene was about 500 times longer than that in UAMD, which is a brute force MD simulation neglecting only hydrogen atoms.

We apply the UEF algorithm to the KG model~\cite{KG1990} and investigate the rheological properties of typical elongational flows, such as uniaxial, biaxial and planar elongational flows. Uniaxial and biaxial elongational flows have not been investigated through conventional methods due to the technical difficulties discussed in a later section, while 
planar elongational flows of linear polymer melts~\cite{Daivis2003,Daivis2007} and branched polymer melts~\cite{Hajizadeh2014} have been investigated by KGMD (KGMD is not described by Langevin dynamics but by Newtonian dynamics) using KRBCs.

We expect that the viscosity growth curves in elongational flows are proportional to the linear viscosity growth curve $\eta_0(t)$ in the linear strain region as shown in the experimental observations~\cite{LS1989}.
The linear viscosity growth curve is obtained from the time integral of the relaxation modulus $G(t)$ as
\begin{equation}
 \eta_0(t)=\int_0^{t} G(t') dt', \label{eq.lin.visc}
\end{equation}
where $G(t)$ is defined in Appendix \ref{app.lin}.
From linear viscoelastic theory~\cite{Larson1999}, the uniaxial elongational viscosity $\eta_{\rm u}(t)$, biaxial elongational viscosity $\eta_{\rm b}(t)$, and planar elongational viscosity $\eta_{\rm p}(t)$ in a linear (or small) strain region correspond to $3\eta_0(t)$, $6\eta_0(t)$, and $4\eta_0(t)$, respectively.
When the strain or strain rate is high, the melts show nonlinear behavior deviating from the linear viscosity growth curves. This is called strain hardening (or strain softening), where the viscosity increases (or decreases) with increasing strain.
Applying the UEF method to the polymer melts, we investigate the linear viscosity and the nonlinear viscosity in elongational flows.
We expect that the UEF method will enable us to obtain the steady-state viscosity at the late stage.

When we handle long polymer chains with the UEF algorithm, 
the finite system size may affect the simulation results because the elongated polymer chains in elongational flows are folded through the periodic boundary conditions (PBCs). These chains, which are elongated but folded through the PBCs, might correlate with themselves when the system size is small. 
Therefore, we need to check the finite size dependency.

The second section shows the simulation methods. The third section shows the simulation results and discussion. We have obtained the viscosity growth curves and the steady-state viscosities in the typical elongational flows. The final section summarizes this work.

\section{Methods}

To investigate the rheological properties of elongational flows,
we use the KG model~\cite{KG1990}.
The system consists of $M$ linear polymer chains, where a single polymer chain is represented by $N$ beads connected by $N-1$ springs.
Each bead has the repulsive LJ potential and the finite-extensible-nonlinear-elastic (FENE) potential:
\begin{align}
 U_{\rm LJ}(r)&=4\epsilon \left\{
\left(
\frac{\sigma}{r}
\right)^{12}
-
\left(
\frac{\sigma}{r}
\right)^{6}
+
\frac{1}{4}
\right\}
, \quad (r < 2^{1/6}\sigma) \\
U_{\rm FENE}(r)&=
-\frac{k R_0^2} {2} \ln \left\{
1 - \left(\frac{r}{R_0}\right)^2
\right\}, \quad (r<R_0)
\end{align}
where $R_0$ is the maximum length of the spring and $k$ is the spring constant.
The LJ energy $\epsilon$, the LJ radius $\sigma$ and the LJ mass $m$ are set to unity.
The elementary bond length relaxation time $\tau=\sigma\sqrt{m/\epsilon}$ corresponds to the unit of time and the unit of stress is $\epsilon/\sigma^3$.
Each LJ particle moves according to the Langevin dynamics equation
\begin{equation}
 m \frac{d^2 \vec{r}_i}{d t^2} = -\sum_{j \ne i}\nabla U(r_{ij}) - \zeta \vec{v}_i + \vec{R}_i(t),
\label{eq.langevin}
\end{equation}
where $\vec{r}_i$ and $\vec{v}_i$ are the position and velocity of the $i$-th particle, respectively,
$r_{ij}=|\vec{r}_j-\vec{r}_i|$,
$\zeta$ is the friction constant, and $\vec{R}_i(t)$ is the random force 
acting on the $i$-th particle.
The origin of the random force
is the vibrational motions of molecules smaller than the LJ particle
which consists of (or corresponds to) several monomer units.
The random force is assumed to be Gaussian with a mean of zero and variance given by
\begin{equation}
 \langle \vec{R}_i(t)\cdot\vec{R}_j(t') \rangle = 6 \zeta k_{\rm B} T \delta_{ij}\delta (t-t'),
\end{equation}
where $k_{\rm B}$ is Boltzmann's constant, and $T$ is the temperature.
The rheological properties are reflected in the stress tensor~\cite{Thompson2009}
\begin{align}
 \sigma_{\alpha \beta}=\frac{-1}{V}
\left(
\sum_{i}^{N_{\rm tot}} m v_i^{\alpha} v_i^{\beta}
+\sum_{i}^{N_{\rm tot}'}  r_i^{\alpha} F_i^{\beta}
\right), \quad (\alpha, \beta=x, y, z) \label{eq.stress}
\end{align}
where $V$ is the system volume, $N_{\rm tot}=MN$ represents the total number of LJ particles in the system, $N_{\rm tot}'$ includes the periodic image particles, and 
$\vec{F}_i$ is the force acting on the $i$-th particle represented in the right side of Eq. \eqref{eq.langevin}.

The values of the parameters summarized in Table~\ref{table.parameter} were determined so that the bead-spring chains cannot cross each other. 
To investigate a weakly entangled linear polymer melt,
$M=1000$ linear chains with $N=100$ beads were randomly generated in a system and the mass density $\rho=mMN/V$ was set to 0.85. 
The size of the unit cell $L=(mMN/\rho)^{1/3}$ is $49.0$.
Equilibration was carried out for a sufficiently longer time than the longest relaxation time $\tau_1$ which was determined from 
the autocorrelation function of the 1st Rouse mode, defined in Appendix~\ref{app.rma}.
The number of entanglement points per chain $Z$ is less than $3$~\cite{KG1990}.

\begin{table}[h]
\caption{Constant parameters and their units.}
\label{table.parameter}
\begin{tabular}{ccc}
\hline
variable & value & unit \\
\hline
 $R_0$ & 1.5 &$[\sigma]$\\
 $k$   & 30 & $[\epsilon/\sigma^2]$ \\
 $k_{\rm B}T$ & 1.0 &$[\epsilon]$\\
 $\zeta$ & 0.5 &$[m / \tau]$\\
 $\rho$ & 0.85 &$[m/\sigma^3]$\\
 $\Delta t$ & 0.01 & $[\tau]$ \\
 $M$ & 1000 & $[ \,\cdot\, ]$ \\
 $N$ & 100  & $[ \,\cdot\, ]$ \\
 $L$ & 49.0  & $[\sigma]$ \\
 
\hline
\end{tabular}
\end{table}

Applying a flow field to the system, the $i$-th particle moves according to the following SLLOD equations~\cite{Evans}:
\begin{equation}
 \frac{d \vec{r}_i}{d t} = \vec{v}_i+\tensor{\kappa} \cdot \vec{r}_i, \quad \frac{d \vec{v}_i}{d t} = \frac{\vec{F}_i}{m}-\tensor{\kappa}^{T} \cdot \vec{v}_i,\label{eq.deform}
\end{equation}
where 
$\tensor{\kappa}=(\nabla \vec{v})^{T}$ is the velocity gradient tensor, and $\vec{v}$ is the external velocity field. 
When we consider a uniaxial elongation with a constant elongational rate $\dot{\epsilon}$, 
\begin{equation}
 \tensor{\kappa}={\rm diag}(-\dot{\epsilon}/2,-\dot{\epsilon}/2,\dot{\epsilon}),
\end{equation}
any position $\vec{r}$ fixed in the system after an interval of time $t$ in the uniaxial elongational flow is
\begin{equation}
 \vec{r}(t)={\rm diag}
(\exp(-\dot{\epsilon}t/2), \exp(-\dot{\epsilon}t/2), \exp(\dot{\epsilon}t)) \cdot \vec{r}(0).
\label{eq.deform.result}
\end{equation}
To satisfy the PBCs, the unit cell is also deformed in accordance with Eq. \eqref{eq.deform.result}. The system then grows exponentially in time.
When $t \gg 1/\dot{\epsilon}$, the system becomes very long along the $z$-axis and narrow in the $xy$-plane, and then the molecules correlate to themselves beyond the periodic boundaries.
Therefore, the simulation in the conventional elongation by the RPS method breaks down at a finite strain that depends on the volume of the system.
From the strain after time interval $t$, we can estimate the size of the rectangular parallelepiped unit cell: $L_z(t)=L_0 \exp(\epsilon)$, $L_x(t)=L_y(t)=L_0 \exp(-\epsilon/2)$, where $\epsilon=\dot{\epsilon}t$ and $L_0$ is the side length of the initial cubic cell.
To achieve a steady state in uniaxial elongational flows, $\epsilon \ge 4.5$ is required from the experimental observation~\cite{LS1989};
that is, $L_z \ge 90.0 L_0$ and $L_x =L_y \le 0.1054 L_0$ when $\epsilon \ge 4.5$.
The major axis $L_z$ becomes 90 times larger than $L_0$, and 
the minor axes $L_x$ and $L_y$ are approximately one-tenth of $L_0$.
Therefore, for the RPS method, the volume of the system that is at least one hundred times larger than the stable volume of the system is needed.
(When we start with $L_x(0)=L_y(0)=10 L_0$ and $L_z(0)=L_0$ where the volume of the system is 100 $L_0^3$, we can get the steady state~\cite{Yaoita2011}.)

The UEF method~\cite{NR2016} can remove the limitation originating in the elongational flow without increasing the computational cost.
At the initial state, the basis vectors of the unit cell $\vec{b}_i^0 \, (i=1,2,3)$ are set not to be parallel to the elongational axes $\vec{e}_{\alpha} \, (\alpha=x,y,z)$. 
The initial basis vectors of the unit cell $\vec{b}_i^0$ are the eigenvectors of the matrix of the automorphisms~\cite{Dobson2014,Hunt2015}. 
The basis vectors $\vec{b}_i^0$ are deformed in the uniaxial elongational flow field in accordance with Eq. \eqref{eq.deform.result}.
Being different from the conventional RPS method, the deformed basis vectors $\vec{b}_i(t)$ can be transformed back into the original cubic configuration 
when the basis vectors $\vec{b}_i(t)$ correspond to the reproducible lattice:
\begin{align}
\vec{b}_i(t)&=\tensor{\Lambda} \cdot \vec{b}_i^0 \label{eq.repro.lattice1}\\
&=M_{i1}\vec{b}_1^0 + M_{i2}\vec{b}_2^0 + M_{i3}\vec{b}_3^0, \label{eq.repro.lattice2}
\end{align}
where $\tensor{\Lambda}=\exp(\tensor{\kappa} t)$, and $M_{ij}$ are integers.
Equations \eqref{eq.repro.lattice1} and \eqref{eq.repro.lattice2} can be regarded as the eigenvalue equation
\begin{equation}
\tensor{M}\cdot\tensor{V}={\rm diag}(\lambda_1,\lambda_2,\lambda_3)\cdot\tensor{V}, \label{eq.eigen}
\end{equation}
where $\tensor{M}$ is the automorphism matrix composed of the integers $M_{ij}$, $\lambda_i=\exp(\kappa_{ii} t)$ is the eigenvalues, and $\tensor{V}$ is the eigenvector matrix composed of $\vec{b}_i^0$.
Several automorphism matrices for elongational flows are summarized in Appendix~\ref{app.automat}.

\begin{figure}[h]
\centering
\includegraphics[width=.9\hsize]{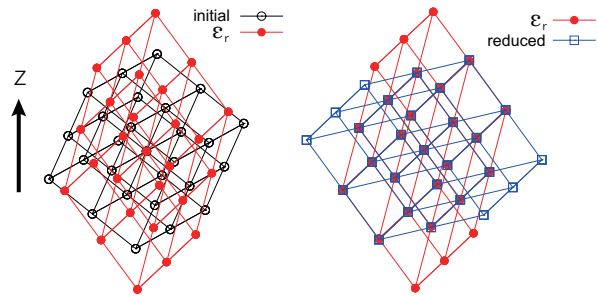} 
 \caption{Deformation (left) and reduction (right) processes of the periodic unit cell in the UEF algorithm. The open circles ``$\circ$'' represent the unit cell at the initial state. The filled circles ``$\bullet$'' represent the unit cell at a certain strain $\epsilon_{\rm r}$ under uniaxial elongation. The squares ``$\Box$'' represent the unit cell after lattice reduction.}
 \label{fig.lattice}  

\end{figure}

Once we have found the initial basis vectors $\vec{b}_i^0$ for elongational flows,
we can apply elongational deformation repeatedly to the system.
For example, 
we assume uniaxial elongation along the $z$-axis.
The left figure shown in Fig.~\ref{fig.lattice} represents the initial state of the unit cell $\vec{b}_i^0$ and the deformed unit cell $\vec{b}_i(t)$ at a finite strain $\epsilon_{\rm r}=\dot{\epsilon}t$. When the strain $\epsilon_{\rm r}$ matches the generalized KRBCs~\cite{Dobson2014,Hunt2015}, that is, when $\vec{b}_i(t)$ corresponds to the reproducible lattice, we can switch the squashed unit cell to the cubic unit cell by shearing in a certain plane as shown in the right figure in Fig.~\ref{fig.lattice}, similar to the shear deformation.
To reduce the lattice, Semaev's algorithm~\cite{Semaev} (see Appendix~\ref{app.semaev}) has been implemented in the UEF method.

The details of the UEF algorithm are found in Refs.~\cite{Dobson2014,Hunt2015,NR2016} and the UEF method is now available as the USER-UEF package in LAMMPS~\cite{LAMMPS}. However, this method is not applicable to Langevin dynamics as it is. We have extended their code~\cite{NR2016} to make it applicable to Langevin dynamics~\cite{UEFEX}. The extension is simple. 
In the Langevin dynamics, the random force $\vec{R}_i(t)$ and the friction force $-\zeta \vec{v}_i$ work as the thermostat. 
The Langevin dynamics itself is treated in the framework of the micro canonical MD.
Because the original UEF code does not support the micro canonical MD with the random force and the friction force, we added this part. 

In the next section, we investigate uniaxial elongational flow of the weakly entangled polymer melt, and then discuss the finite size effect. Furthermore, biaxial elongational flow and planar elongational flow are also analyzed using our extended UEF code.

\section{Results and Discussion}

In this section, we will show the numerical results obtained by our extended UEF code.
In the first subsection, we investigate uniaxial elongational flow and compare the results obtained with the RPS method and the UEF method.
We then focus on the finite size effect.
Finally, we investigate biaxial elongational flow and planar elongational flow.

\subsection{Uniaxial elongational flow: Comparison between the RPS method and the UEF method\label{subsec1}}

In this subsection, we investigate the weakly entangled polymer melts under the uniaxial elongational flow $\tensor{\kappa}={\rm diag}(-\dot{\epsilon}/2, -\dot{\epsilon}/2, \dot{\epsilon})$ and compare the conventional RPS method and the UEF method.
We observe the viscosity growth curves of the uniaxial elongational viscosity
$\eta_{\rm u}(t)=\sigma_{\rm u}(t)/\dot{\epsilon}$, where $\sigma_{\rm u}=\sigma_{zz}-(\sigma_{xx}+\sigma_{yy})/2$. 
To reduce the noise in the data, the moving time averaging $\bar{A}(t)=\int_{t-\Delta}^{t+\Delta} A(t) dt / \int_{t-\Delta}^{t+\Delta}dt$, where $\Delta=10^k$ and $k$ is the maximum integer less than or equal to $\log_{10} t -1$, was carried out for the viscosity growth curves. 
Data in the early stage, $t < 0.01/\dot{\epsilon}$, were omitted for clarity.
The linear viscosity growth curve $\eta_0(t)$ was determined from the relaxation modulus $G(t)$, which is summarized in Appendix~\ref{app.lin}.

\begin{figure}[h]
\centering
\includegraphics[width=.9\hsize]{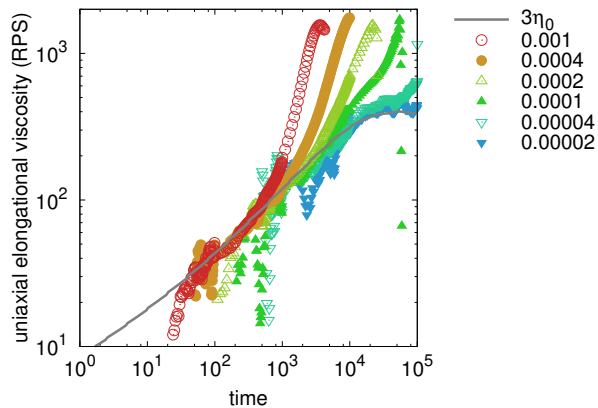}
\caption{Time-dependent uniaxial elongational viscosity $\eta_{\rm u}(t)$ obtained by the conventional RPS method. The number of polymer chains $M$ is 1000. The solid line shows the linear viscosity growth curve $3 \eta_0(t)$, and the symbols represent the elongational rate $\dot{\epsilon}$ from 0.00002 to 0.001.}
 \label{fig.uni-conv}  
\end{figure}

At first, we tested the conventional RPS method to obtain the viscosity growth curve
under uniaxial elongational flow. 
Figure~\ref{fig.uni-conv} shows the viscosity growth curves of the uniaxial elongational viscosity obtained by the RPS method~\cite{Heyes1985}.
As shown in Fig.~\ref{fig.uni-conv}, 
linear viscosity in the linear strain region and strain hardening behaviors in the high-strain region were observed.
However, each viscosity growth curve ends at a finite strain when $\dot{\epsilon} \ge 0.00004$. 
At the end, the simulation breaks down when the size of the cell is comparable to the LJ particle size.
We cannot obtain the steady-state viscosity from the RPS method for high strain rates.

\begin{figure}[h]
\centering
 \includegraphics[width=.9\hsize]{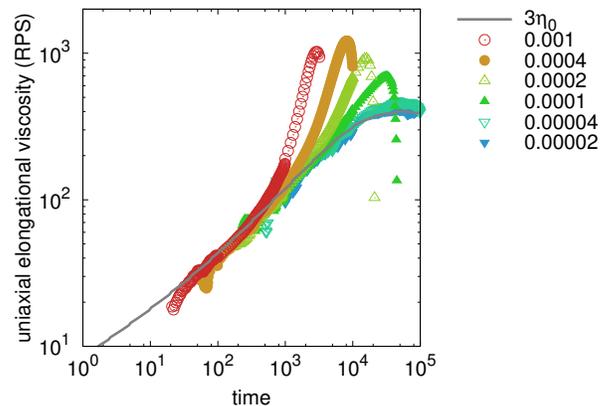}
 \caption{Time-dependent uniaxial elongational viscosity $\eta_{\rm u}(t)$ obtained by the conventional RPS method. The number of polymer chains $M$ is 10000. The solid line shows the linear viscosity growth curve $3 \eta_0(t)$, and the symbols represent the elongational rate $\dot{\epsilon}$ from 0.00002 to 0.001.}
 \label{fig.uni-conv2}  
\end{figure}

By increasing the number of chains $M$, it is believed that the finite size effect can, in general, be decreased.
Figure~\ref{fig.uni-conv2} shows the case of $M=10000$, a system that is ten times larger than the reference system. 
The viscosity growth curves show steady states when $\dot{\epsilon}<0.0001$.
When $\dot{\epsilon}\ge 0.0001$, however, the viscosity growth curve shows a peak at a finite strain $\epsilon \approx 3$ and then rapidly decreases to zero. 
This strain $\epsilon \approx 3$ is smaller than the experimental one $\epsilon \ge 4.5$ where the steady state was observed~\cite{LS1989}.
The anisotropy of the unit cell starts to affect the viscosity at this peak.
To avoid the finite size effect, we need a large number of chains, that is, $M>10000$, which would require enormous computational resources.

\begin{figure}[h]
\centering
 \includegraphics[width=.9\hsize]{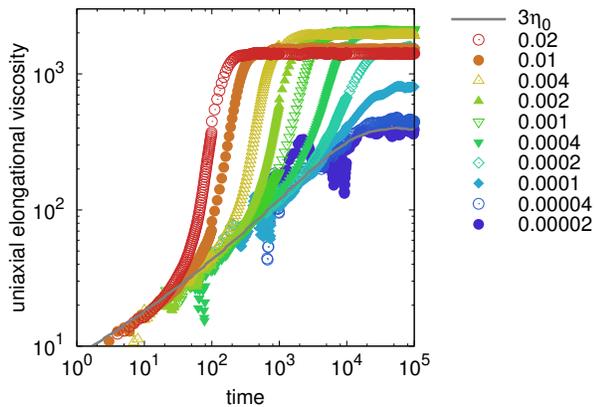}
 \caption{Time-dependent uniaxial elongational viscosity $\eta_{\rm u}(t,\dot{\epsilon})$ obtained by the UEF method.
The number of polymer chains $M$ is 1000. 
The solid line represents the linear viscosity growth curve $3\eta_0(t)$, and the symbols represent the elongational rates $\dot{\epsilon}$ from 0.00002 to 0.02.}
 \label{fig.uni}  
\end{figure}

Next, we investigated uniaxial elongational flow using the UEF method.
The growth curves of the uniaxial elongational viscosity $\eta_{\rm u}(t, \dot{\epsilon})$ are shown in Fig.~\ref{fig.uni}.
Each viscosity growth curve corresponds to the linear viscosity growth curve $3\eta_0(t)$ as long as the strain $\epsilon=\dot{\epsilon}t$ is less than unity. When the strain $\epsilon$ is larger than unity, strain hardening behavior is observed. When $\epsilon \ge 4.5$, the viscosity reaches the steady state, which is consistent with the experimental observation~\cite{LS1989}.

\begin{figure}[h]
\centering
\includegraphics[width=.9\hsize]{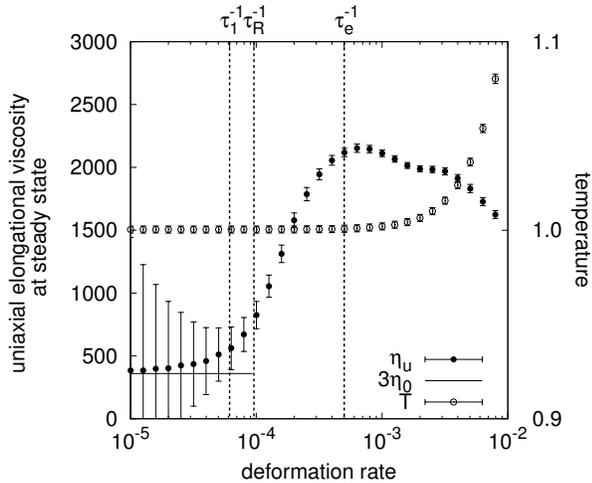}
 \caption{Uniaxial elongational viscosities (the left vertical axis) and temperatures (the right vertical axis) at the steady state plotted against the deformation rate $\dot{\epsilon}$. When $\dot{\epsilon} > 10^{-3}$, the temperature is higher than unity. The number of polymer chains in the system $M$ is $1000$, and the size of the unit cell $L$ is $49.0$. The error bar represents the standard deviation.}
 \label{fig.uni-ss}  
\end{figure}

Figure~\ref{fig.uni-ss} represents the steady-state viscosity of the uniaxial elongational flow 
$\eta_{\rm u}(\dot{\epsilon})$, plotted against the deformation rate $\dot{\epsilon}$.
To obtain the steady-state viscosity, we carried out time averaging in the steady region: 
$\eta_{\rm u}(\dot{\epsilon})=\int_{t_{\rm min}}^{t_{\rm max}} \eta_{\rm u}(t,\dot{\epsilon}) dt \, / \int_{t_{\rm min}}^{t_{\rm max}} dt$, where
$(t_{\rm min}, t_{\rm max})=(10^5, 2\times 10^5)$ for $\dot{\epsilon}\ge 10^{-4}$ and $(t_{\rm min}, t_{\rm max})=(10^6, 2\times 10^6)$ for $\dot{\epsilon}< 10^{-4}$.
The longest relaxation time $\tau_1$ and the Rouse relaxation time $\tau_{\rm R}$ were determined from the autocorrelation function of the 1st Rouse mode, as shown in Appendix~\ref{app.rma}.

The steady-state viscosity shows extensional thickening, where the viscosity increases with 
the deformation rate.
When  
$\dot{\epsilon} \ll 1/\tau_1$,
the relaxation of the polymer chain is faster than the deformation of the flow, and then the steady-state viscosity corresponds to $3\eta_0(\infty)=360$ within the range of statistical error.
When $\dot{\epsilon} \approx 1/\tau_1$,
the steady-state viscosity slightly higher than $3\eta_0(\infty)$.
The steady-state viscosity monotonically increases with $\dot{\epsilon}$ 
when $1/\tau_1<\dot{\epsilon}<0.00063$, where the polymer chains cannot be relaxed at the steady state.
Especially in the region $1/\tau_{\rm R} < \dot{\epsilon} < 1/\tau_{\rm e}$, the viscosity drastically increases with the deformation rate $\dot{\epsilon}$.
The highest value of the steady-state viscosity $\eta_{\rm u}^{\rm Max}$ is $2153$ at $\dot{\epsilon}=0.00063$.
When $\dot{\epsilon}>0.00063$, the steady-state viscosity decreases.
When $\dot{\epsilon} = 0.002$ where the slope of $\eta_{\rm u}(\dot{\epsilon})$ changes discontinuously, the temperature is larger than 1.01.

The typical time for the onset of the reptation motion $\tau_{\rm e}$ has been estimated as $2000$~\cite{KG1990}.
When $\dot{\epsilon}>1/\tau_{\rm e}=5.0\times 10^{-4}$,
the entanglements among chains are not generated by the reptation motion.
In this high-deformation-rate region, the polymer chains orient along the $z$-axis. 
When $\dot{\epsilon}>0.002$, the steady-state viscosity shows plateau. In this region, however, the temperature becomes higher than unity, and the temperature is not controlled. 
In the case of a high-elongational-rate flow,
the heat supplied from the outside is larger than the heat absorbed by the heat bath of the Langevin thermostat.
The Langevin thermostat fails at the high-strain-rate region.
Although the values of the steady-state viscosity when $\dot{\epsilon}>0.002$ exceeds the scope of the application of the Langevin thermostat, we have plotted them for reference.
Even if we use the smaller time width $\Delta t = 0.001$ (which is one tenth the original one in Table.~\ref{table.parameter} ), high-elongational-rate flow also exhibited the same results: the temperature was not controlled by the Langevin thermostat, as shown in Appendix~\ref{app.temp}.
This temperature growth comes from the excluded volume effect in the high-deformation-rate flow, discussed in Appendix~\ref{app.bond}.

When $\dot{\epsilon}<1/\tau_1$, the error bar in the steady-state viscosity increases with decreasing $\dot{\epsilon}$.
These errors are caused by the thermal noise.
By increasing the number of polymer chains $M$, we can obtain more accurate data in the low deformation rate region.

\begin{figure}[h]
\centering
\includegraphics[width=.9\hsize]{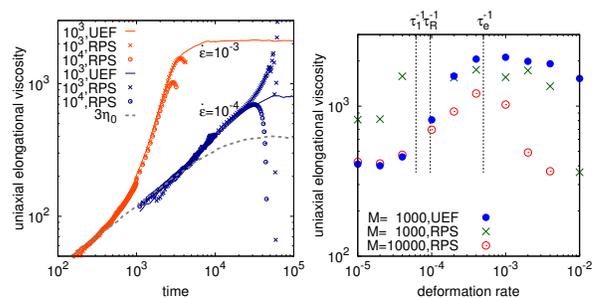} 
 \caption{Comparison between the results of the UEF method and the RPS method; 
time-dependency (left) and steady-state (right).}
 \label{fig.uni-compare}  
\end{figure}

Figure \ref{fig.uni-compare} compares the results obtained by the UEF method and the RPS method.
For the short time period in Fig. \ref{fig.uni-compare}(left), both methods show good agreements.
The steady-state viscosity obtained by the UEF method and the peak value of the viscosity obtained by the RPS method are compared in Fig. \ref{fig.uni-compare}(right).
The error bars shown in Fig. \ref{fig.uni-ss} are omitted for clarity.
Although we know that the peak viscosity obtained by the RPS method with $M=1000$ chains is not accurate, it is displayed here for reference.
When $\dot{\epsilon}<1/\tau_1$, we can see good agreement between the steady-state viscosity obtained by the UEF method and the peak viscosity obtained by the RPS method with $M=10000$ chains.
When $\dot{\epsilon}$ is increased, the deviation is apparent between the steady-state viscosity obtained by the UEF method and the peak viscosity obtained by the RPS method with $M=10000$ chains. 
The peak values of the viscosity appear at the finite strain $\epsilon\approx 3$ in the RPS method with $M=10000$. 
This strain is smaller than the value of the experimentally observed steady state~\cite{LS1989}.
The conventional RPS method underestimates the steady-state viscosity due to the finite size effect.
Moreover, the RPS method requires a large system with $M>10000$, whereas the UEF method uses only $M=1000$ chains to reproduce the steady state.

The UEF method succeeded in obtaining the steady-state viscosity, whereas the RPS method failed to do so.
In the next subsection, we will estimate an efficient system size to investigate the rheological properties of polymer melts with the UEF method.

\subsection{Finite size effect in the UEF method}

In this subsection, we focus on the system size in the UEF method.
A polymer chain in the unit cell with size $L$ under PBCs is separated by $L$ from itself.
When this size $L$ is smaller than the polymer size, the polymer chain can correlate with itself through PBCs.
The size of the polymer is approximately represented by the radius of gyration $R_{\rm g}$ defined as follows:
\begin{equation}
 R_{\rm g}^2 = \langle \frac{1}{N}\sum_{i=1}^N (\vec{r}_i - \vec{r}_{\rm cm})^2 \rangle,
\end{equation}
where $\vec{r}_{\rm cm}=(1/N)\sum_{i=1}^N \vec{r}_i$ is the center of mass of the chain.
At the equilibrium state with $M=1000$, $R_{\rm g}=5.1 (\pm 1.2)$.
The system size $L$ should be at least larger than $2R_{\rm g}$ to prevent overlapping with itself.
The stress tensor~\eqref{eq.stress} is the statistical variable with variance proportional to $1/{N_{\rm tot}}$.
To decrease the variance, we need to handle the sufficiently large number of particles $N_{\rm tot}$ in the system.

\begin{figure}[h]
\centering
 \includegraphics[width=.9\hsize]{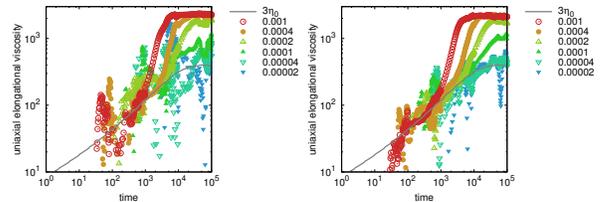} 
 \caption{Time-dependent uniaxial elongational viscosity $\eta_{\rm u}(t,\dot{\epsilon})$ obtained by the UEF method.
The numbers of polymer chains $M$ are 10 (left) and 100 (right). 
The solid line represents the linear viscosity growth curve $3\eta_0(t)$,
and the symbols represent the elongational rates $\dot{\epsilon}$ from 0.00002 to 0.001.}
 \label{fig.uni-10}  
\end{figure}

Figure~\ref{fig.uni-10} shows the viscosity growth curves in uniaxial elongational flows with $M=10$ (left) and $M=100$ (right).
The system size $L$ is 10.5 when $M=10$, and $L=22.7$ when $M=100$.
In both cases, it is difficult to find the correspondence with the linear viscosity growth curve $3\eta_0(t)$  in the small-strain regime when $\dot{\epsilon} \le 0.00004 (<1/\tau_1)$ due to the thermal noise.
On the other hand, the steady-state viscosity shows the correspondence between these cases within the error bar when $\dot{\epsilon} \ge 0.0002$.
The steady-state viscosity in the intermediate-strain-rate region, $1/\tau_1 < \dot{\epsilon}<1/\tau_{\rm e}$, can be estimated when $M=100$, whereas it cannot when $M=10$.
To investigate the flow when $\dot{\epsilon}>1/\tau_1$, we need $M \ge 100$ at least from the rheological standpoint.
In this work, we do not focus on individual molecular motions but on the rheological properties.
Because the rheological properties are insensitive to individual molecular motions,
it is difficult to observe self-correlation between a chain and its PBC image.
If self-correlation occurs, the dynamics of the self-correlating chain will be different from the other dynamics.
Further investigations for the molecular motions in the UEF framework are under way.

\subsection{Biaxial elongational flow and planar elongational flow}

In this subsection, 
we investigate the biaxial elongational flow $\tensor{\kappa}={\rm diag}(\dot{\epsilon}, \dot{\epsilon}, -2\dot{\epsilon})$ 
and the planar elongational flow
$\tensor{\kappa}={\rm diag}(\dot{\epsilon}, -\dot{\epsilon}, 0)$ 
using the UEF method.
As with uniaxial elongational flow,
we observe the biaxial elongational viscosity $\eta_{\rm b}(t, \dot{\epsilon})=\sigma_{\rm b}(t)/\dot{\epsilon}$ and the planar elongational viscosity
$\eta_{\rm p}(t, \dot{\epsilon})=\sigma_{\rm p}(t)/\dot{\epsilon}$,
where $\sigma_{\rm b}=-\sigma_{\rm u}$ and $\sigma_{\rm p}=\sigma_{xx}-\sigma_{yy}$ are the first normal stress difference of the biaxial and planar elongational flows, respectively. 

\begin{figure}[h]
\centering
 \includegraphics[width=.9\hsize]{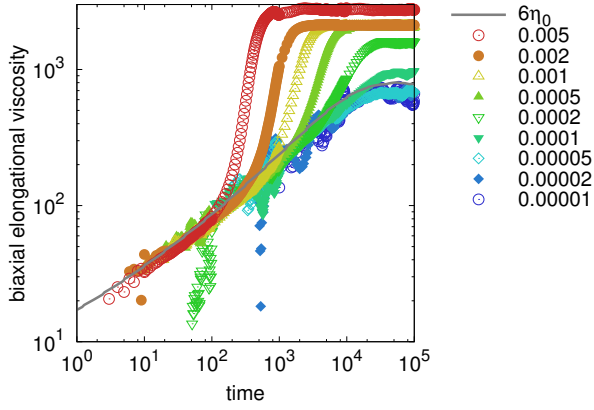}
 \caption{Time-dependent biaxial elongational viscosity $\eta_{\rm b}(t,\dot{\epsilon})$ obtained by the UEF method. 
The solid line represents the linear viscosity growth curve $6\eta_0(t)$,
and the symbols represent the elongational rate $\dot{\epsilon}$ from 0.00001 to 0.005.}
 \label{fig.bi}  
\end{figure}

\begin{figure}[h]
\centering
\includegraphics[width=.9\hsize]{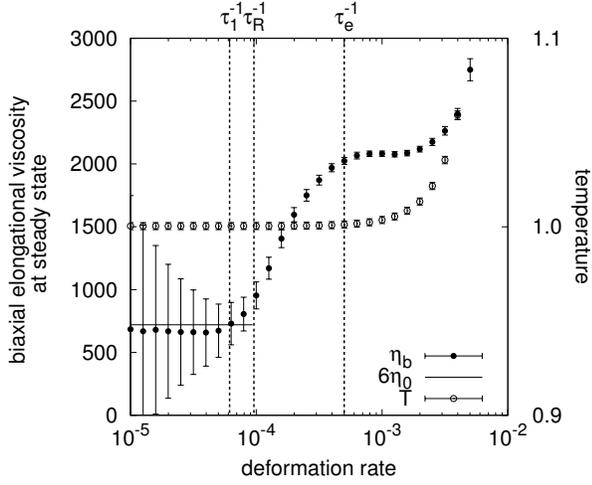}
 \caption{Biaxial elongational viscosities (the left vertical axis) and temperatures (the right vertical axis) at the steady state plotted against the deformation rate $\dot{\epsilon}$.
The number of polymer chains in the system $M$ is $1000$, and the size of the unit cell $L$ is $49.0$. The error bar represents the standard deviation.}
 \label{fig.bi-ss}  
\end{figure}

The growth curves of the biaxial elongational viscosity $\eta_{\rm b}(t, \dot{\epsilon})$ are shown in Fig.~\ref{fig.bi}. In the small-strain region of the biaxial elongational flow, the viscosity growth curves correspond to $6\eta_0(t)$. As the strain increases, strain hardening behavior is observed and then the viscosity reaches the steady-state value as well as uniaxial elongational flow.
Figure~\ref{fig.bi-ss} shows the steady-state viscosity of the biaxial elongational flow $\eta_{\rm b}(\dot{\epsilon})$. When $\dot{\epsilon}<1/\tau_1$, the steady-state viscosity is slightly smaller than $6\eta_0(\infty)=720$. 
The minimum value $\eta_{\rm b}^{\rm Min}$ is $658$ at $\dot{\epsilon}=0.000040$.
When $1/\tau_1 < \dot{\epsilon}<0.001$, the steady-state viscosity monotonically increases and then shows plateau.
When $\dot{\epsilon}=0.002$, the temperature exceeds 1.01.
The maximum value of the steady-state viscosity $\eta_{\rm b}^{\rm Max}$ within $T<1.01$ is $2082$ at $\dot{\epsilon}=0.001$.

The maximum value of the steady-state viscosity in the biaxial elongational flow $\eta_{\rm b}^{\rm Max}$ is smaller than that in the uniaxial elongational flow $\eta_{\rm u}^{\rm Max}$.
These elongational flows cause the anisotropy in the orientation of polymer chains at the steady state.
While the polymer chains orient along the $z$-axis in the uniaxial elongational flow, the polymer chains in the biaxial elongational flow are stretched in the $xy$-plane.
The orientational order in the uniaxial elongational flow is higher than that in the biaxial elongational flow, resulting in the difference between these maximum values of the steady-state viscosity.

The minimum value of the steady-state viscosity in the biaxial elongational flow $\eta_{\rm b}^{\rm Min}$ is smaller than the value of the linear viscosity $6\eta_0(\infty)$, while the uniaxial elongational flow shows the correspondence between the linear viscosity and the minimum value of the steady-state viscosity.
In general, a low Weissenberg number (Weissenberg number $Wi=\dot{\epsilon}\tau_1$) of less than unity is expected to represent the isotropic state of the system, namely, the Newtonian state.
Because the system size grows exponentially in time in elongational flows,
the anisotropy in the orientation of polymer chains exists even in the low-Weissenberg-number flow.

\begin{figure}[h]
\centering
 \includegraphics[width=.9\hsize]{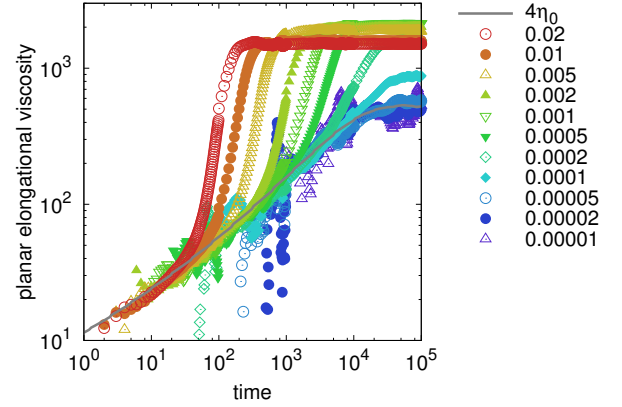}
 \caption{Time-dependent planar elongational viscosity $\eta_{\rm p}(t,\dot{\epsilon})$ obtained by the UEF method. 
The solid line represents the linear viscosity growth curve $4\eta_0(t)$,
and the symbols represent the elongational rate $\dot{\epsilon}$ from 0.00001 to 0.02.}
 \label{fig.pl}  
\end{figure}

\begin{figure}[h]
\centering
\includegraphics[width=.9\hsize]{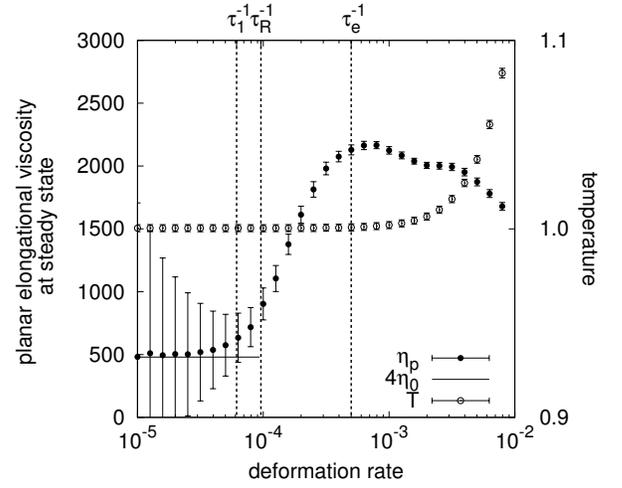}
 \caption{Planar elongational viscosities (the left vertical axis) and temperatures (the right vertical axis) at the steady state plotted against the deformation rate $\dot{\epsilon}$.
The number of polymer chains in the system $M$ is $1000$, and the size of the unit cell $L$ is $49.0$. The error bar represents the standard deviation.}
 \label{fig.pl-ss}  
\end{figure}

Finally, we investigated planar elongational flow using the UEF method.
The growth curves of the planar elongational viscosity $\eta_{\rm p}(t, \dot{\epsilon})$ are shown in Fig.~\ref{fig.pl}. 
In the small-strain region of the planar elongational flow, the viscosity growth curves correspond to $4\eta_0(t)$ and show strain hardening behavior in the high-strain region. 
Figure~\ref{fig.pl-ss} shows the steady-state viscosity of the planar elongational flow $\eta_{\rm p}(\dot{\epsilon})$.
When $\dot{\epsilon} < 1/\tau_1$, the steady-state viscosity in the planar elongational flow corresponds to the linear viscosity $4\eta_0(\infty)=480$ within the error bar.
When $1/\tau_1<\dot{\epsilon}<0.00079$, the steady-state viscosity increases monotonically.
The maximum value of the steady-state viscosity $\eta_{\rm p}^{\rm Max}$ is $2166$ at $\dot{\epsilon}=0.00079$.
When $\dot{\epsilon}>0.002$, the temperature exceeds 1.01.

\begin{figure}[h]
\centering
\includegraphics[width=.9\hsize]{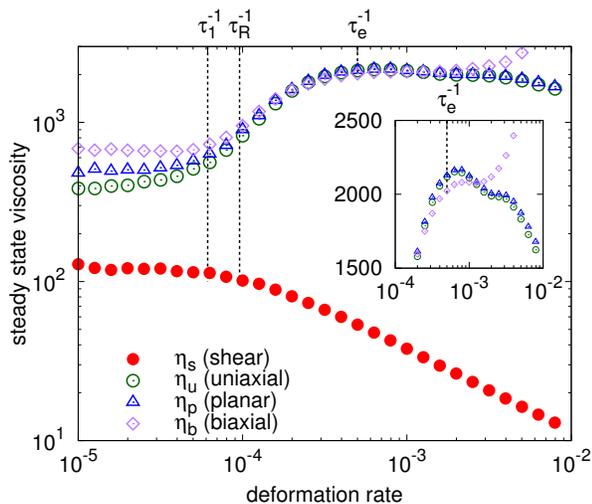}
 \caption{Comparison between the steady-state viscosities of shear flow, uniaxial, planar, and biaxial elongational flows plotted against the deformation rate.}
 \label{fig.ss}  
\end{figure}

Figure \ref{fig.ss} summarizes the steady-state viscosities of the shear flow $\eta_{\rm s}(\dot{\epsilon})$, uniaxial elongational flow $\eta_{\rm u}(\dot{\epsilon})$, planar elongational flow $\eta_{\rm p}(\dot{\epsilon})$, and biaxial elongational flow $\eta_{\rm b}(\dot{\epsilon})$. 
The steady-state viscosity of the shear flow $\eta_{\rm s}(\dot{\gamma})$ is obtained in Appendix \ref{app.lin}.
Error bars are omitted for clarity.
When $\dot{\epsilon}<1/\tau_1$, we found that each viscosity corresponds to the linear viscosity: $\lim_{\dot{\gamma} \to 0}\eta_{\rm s}(\dot{\gamma})=\eta_0(\infty)$,
$\lim_{\dot{\epsilon} \to 0}\eta_{\rm u}(\dot{\epsilon})=3\eta_0(\infty)$, and
$\lim_{\dot{\epsilon} \to 0}\eta_{\rm p}(\dot{\epsilon})=4\eta_0(\infty)$,
while
$\lim_{\dot{\epsilon} \to 0}\eta_{\rm b}(\dot{\epsilon}) < 6\eta_0(\infty)$.
Increasing $\dot{\epsilon}$ brings these three elongational viscosities close.
When $\dot{\epsilon}>1/\tau_{\rm e}$, the uniaxial and planar elongational viscosities show good agreement, whereas the biaxial elongational viscosity shows a different behavior.
This agreement between the uniaxial and planar elongational viscosities in the high-deformation-rate region is in accordance with the theoretical prediction~\cite{LS1989}.
On the steady-state uniaxial elongational viscosity and the steady-state shear viscosity,
our results are consistent with the experimental finding~\cite{LS1989}.

\section{Conclusion}

We investigated the uniaxial, biaxial, and planar elongational flows of the weakly entangled polymer melt with the KG model using the extended UEF method, which we developed.
We succeeded in obtaining the viscosity growth curves and the steady-state viscosities in these elongational flows, although the conventional RPS method has failed to do so for over 20 years.
In the low-deformation-rate region, $\dot{\epsilon}<1/\tau_1$, the steady-state viscosities of the uniaxial and planar elongational flows almost equal $3\eta_0(\infty)$ and $4\eta_0(\infty)$, respectively, although that of the biaxial elongational flow shows a slightly smaller value than $6\eta_0(\infty)$.
In the intermediate-deformation-rate region between $1/\tau_1$ and $1/\tau_{\rm e}$, the steady-state viscosity increases monotonically with $\dot{\epsilon}$.
Especially in the region $1/\tau_{\rm R} < \dot{\epsilon} < 1/\tau_{\rm e}$, the steady-state viscosity drastically increases with $\dot{\epsilon}$.
When $1/\tau_{\rm e}<\dot{\epsilon}$, the steady-state viscosity show a peak in the uniaxial and planar elongational flows, and show plateau in the biaxial elongational flow.
In the high-elongational-rate region, $\dot{\epsilon}>0.002(>1/\tau_{\rm e})$, the temperature was not controlled by the Langevin thermostat.
This temperature growth can be related to viscous heating~\cite{Bird2002}.
The Langevin thermostat and also the other thermostats are based on the equilibrium theory, and they assume the constant temperature. 
Therefore, the local heating process is not considered well within the conventional framework.
Further improvements of the thermostats for non-equilibrium dynamics are needed.

Although we have confirmed that the UEF method has been successful in obtaining the rheological properties of polymer melts, we have found several issues to consider.
The molecules in the UEF framework can correlate with themselves through PBCs.
To disturb the self-correlation, we need to estimate an optimal system size by observing molecular motions.
The further investigations, not only for polymer melts but also for grafted nano particles, are now under way. 
The current UEF method assumes isotropy at the initial state.
If we want to apply an elongation to an anisotropic system, e.g. a lamellar phase in a block copolymer melt, and observe the correlation between the elongational direction and the anisotropy, we need to prepare the initial state carefully.
It is necessary to increase the choices of initial basis vectors $\vec{b}_i^0$.
For this purpose, a search for automorphism matrices is important.

Our numerical results for the case of the weakly entangled polymer melt show extensional thickening, where the steady-state viscosity increases with the deformation rate.
Recently, experiments have shown controversial results for entangled polymer melts and entangled polymer solutions~\cite{Bhattacharjee2002,Bach2003,Sridhar2014,Huang2015}.
Entangled polymer solutions have represented extensional thickening~\cite{Bhattacharjee2002}, while 
entangled polymer melts have represented extensional thinning~\cite{Bach2003}.
Crossover between them has been observed, changing the concentration of polymers in solutions while keeping constant the number of entanglements~\cite{Sridhar2014,Huang2015}.
Ianniruberto has proposed that the extensional thickening is due to the gradual loss of friction-coefficient reduction with decreasing polymer concentration or the entanglement density~\cite{Ianniruberto2012,Yaoita2012,Ianniruberto2015}. 
Although our case is not the entangled polymer solution but the weakly entangled polymer melt with $Z < 3$, extensional thickening has been observed due to the small entanglement density in our system.
We are now conducting analysis of well entangled polymer melts and solutions.
Further reports will be released in the future.

The UEF method is useful for the other CG models for entangled polymer melts~\cite{Mas2001,Sirk2012,UM2012}.
Furthermore, it can extend the range of application of the multiscale simulation which is concurrently solving macroscopic fluid dynamics and microscopic polymer dynamics~\cite{TM2011,TM2013}. 
In particular, the multiscale simulation of fiber spinning~\cite{Sato2017} and the film extruding process will be our future targets. 

\begin{acknowledgments}
TM thanks Prof. J.-I. Takimoto, Prof. T. Taniguchi, and Prof. T. Uneyama for fruitful discussions on elongational flows.
This research was supported by MEXT as ``Exploratory Challenge on Post-K computer'' (Challenge of Basic Science - Exploring Extremes through Multi-Physics and Multi-Scale Simulations) and JSPS KAKENHI Grant Number 15K17733.
This research used the computational resources of the K computer provided by the RIKEN Advanced Institute for Computational Science through the HPCI System Research project (Project ID:hp160267/hp170236/hp180116) and the facilities of the Supercomputer Center, the Institute for Solid State Physics, the University of Tokyo.
The model polymers were generated using OCTA-COGNAC~\cite{OCTA-COGNAC} and the productive runs were performed by LAMMPS~\cite{LAMMPS}.
We would like to thank Editage (www.editage.jp) for English language editing.

\end{acknowledgments}

\appendix

\section{Linear viscoelasticity and shear viscosity \label{app.lin}}

\begin{figure}[h]
\centering
 \includegraphics[width=.9\hsize]{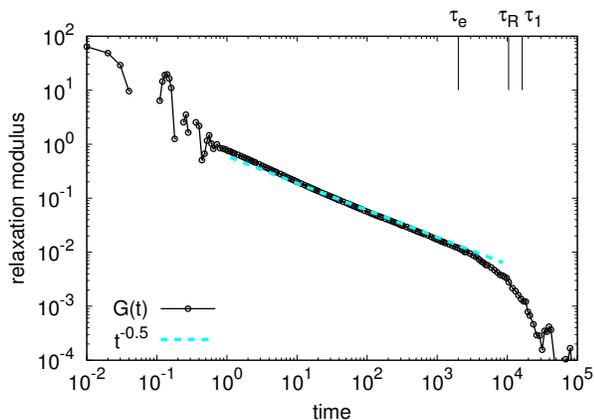}
 \caption{Relaxation modulus $G(t)$ and the Rouse relaxation behavior $G(t)\sim t^{-1/2}$. 
The characteristic times $\tau_{\rm e}$, $\tau_{\rm R}$ and $\tau_1$ are $2000$, $10433$ and $16308$, respectively, obtained in Appendix~\ref{app.rma}.
}
 \label{fig.gt}  
\end{figure}

The relaxation modulus $G(t)$ is obtained from the auto-correlation function of the shear stress $\sigma_{xy}$. To improve accuracy, we average the auto-correlation functions over the possible directions:
\begin{equation}
\begin{split}
  G(t)&=\frac{V}{6 k_{\rm B}T}[
\langle \sigma_{xy}(t)\sigma_{xy}(0)\rangle
+\langle \sigma_{yz}(t)\sigma_{yz}(0)\rangle
\\
&\quad\quad\quad\quad\quad+\langle \sigma_{zx}(t)\sigma_{zx}(0)\rangle
] \\
&+
\frac{V}{24 k_{\rm B}T}
[
\langle N_{xy}(t)N_{xy}(0)\rangle
+\langle N_{yz}(t)N_{yz}(0)\rangle
\\
&\quad\quad\quad\quad\quad+\langle N_{zx}(t)N_{zx}(0)\rangle
],
\end{split}
\end{equation}
where $N_{\alpha \beta}=\sigma_{\alpha\alpha}-\sigma_{\beta\beta}$. Here, we assume linear viscoelasticity and isotropy for the equilibrium system.
The auto-correlation function $\langle \sigma(t)\sigma(0) \rangle$ was obtained by using the multiple-tau method~\cite{Ramirez2010}.
Figure~\ref{fig.gt} shows the relaxation modulus. The Rouse relaxation behavior $G(t)\sim t^{-1/2}$ is found in the middle time range.

\begin{figure}[h]
\centering
 \includegraphics[width=.9\hsize]{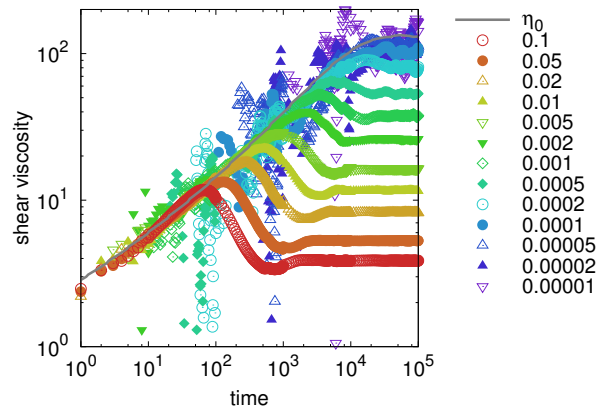}
 \caption{Time-dependent shear viscosity $\eta_{\rm s}(t,\dot{\gamma})=\sigma_{xy}(t)/\dot{\gamma}$ under constant shear flows with shear rates $\dot{\gamma}$ from 0.00001 to 0.1. The solid line represents the linear viscosity growth curve $\eta_0(t)$ obtained from the relaxation modulus $G(t)$.}
 \label{fig.sh}  
\end{figure}

Integrating $G(t)$ as shown in Eq. \eqref{eq.lin.visc}, we obtain the linear viscosity growth curve $\eta_0(t)$.
In order to check the linear viscosity growth curve $\eta_0(t)$, we have compared it with the shear viscosity growth curves $\eta_{\rm s}(t,\dot{\gamma})=\sigma_{\rm s}(t)/\dot{\gamma}$, where $\sigma_{\rm s}=\sigma_{xy}$ and $\dot{\gamma}=\kappa_{xy}$. 
To apply shear flow to the system, we used the Lagrangian rhomboid boundary conditions (LRBCs)~\cite{Evans1979,HE1994,HT2003}.
The LRBCs are mathematically equivalent to the Lees--Edwards boundary conditions (LEBCs)~\cite{LE1972}.
Furthermore, the LRBCs are more suitable for parallel computing than the LEBCs~\cite{HE1994}.
We can confirm that the shear viscosity growth curve $\eta_{\rm s}(t)$ within a small strain $\gamma=\dot{\gamma} t \approx 1$ corresponds to the linear viscosity growth curve $\eta_0(t)$ as shown in Fig.~\ref{fig.sh}.

\begin{figure}[h]
\centering
\includegraphics[width=.9\hsize]{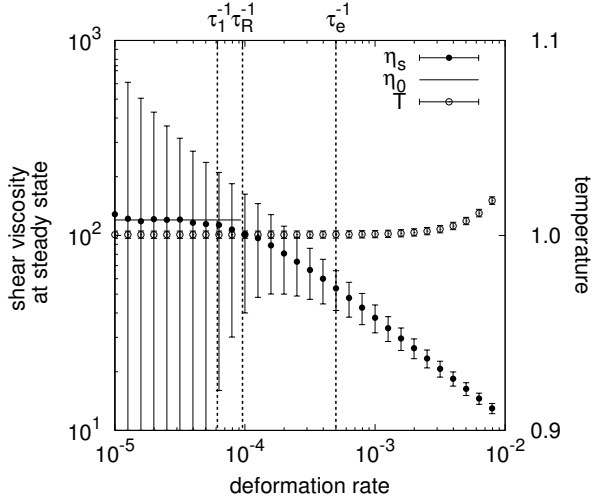}
 \caption{Shear viscosities (left vertical axis) and temperatures (right vertical axis) at the steady state. The horizontal axis represents 
the shear rate $\dot{\gamma}$. 
The horizontal line shown in the small shear rate represents the zero shear viscosity $\eta_0=\eta_0(\infty)$. 
The number of polymer chains in the system $M$ is $1000$. The error bar represents the standard deviation.}
 \label{fig.sh-ss}  
\end{figure}

The steady-state viscosity was obtained by time-averaging over the steady region, $10^5 < t < 2\times 10^5$ for $\dot{\gamma} \ge 10^{-4}$ and $10^6 < t < 2\times 10^6$ for $\dot{\gamma}<10^{-4}$.
Figure~\ref{fig.sh-ss} shows the shear viscosity $\eta_{\rm s}(\dot{\gamma})$ and the temperature $T$ at the steady state plotted against 
the shear rate $\dot{\gamma}$.
When $\dot{\gamma}<\tau_1^{-1}$, the steady-state viscosity corresponds to the zero-shear viscosity $\eta_0(\infty)=120$ within the error bar.
When $\dot{\gamma}>\tau_1^{-1}$, the steady-state viscosity is smaller than $\eta_0(\infty)$, representing shear-thinning behavior. 
When $\dot{\gamma}>0.005$, the temperature is larger than unity.
The obtained steady-state viscosity shown in Fig. \ref{fig.sh-ss} is consistent with previous works~\cite{Aoyagi2000,Kroger2000}.

\section{Rouse mode relaxation and relaxation times}\label{app.rma}

In this appendix, we determine the characteristic relaxation times of weakly entangled polymer chain with $N=100$.
The longest relaxation time $\tau_1$ and the Rouse relaxation time $\tau_{\rm R}$ are obtained from the Rouse mode relaxation~\cite{KG1990,RMA2017}.
The $p$-th Rouse mode is
\begin{equation}
 \vec{X}_p(t)=\sqrt{\frac{2}{N}}\sum_{i=1}^{N}\vec{r}_i(t)\cos\left(\frac{p\pi}{N}\left(i-\frac{1}{2}\right)\right).
\end{equation}
The normalized Rouse mode autocorrelation function of polymer melt is well described by the stretched exponential Kohlrausch-Williams-Watts (KWW) function,
\begin{equation}
 \frac{\langle \vec{X}_p(t+t')\cdot\vec{X}_p(t') \rangle}
{\langle \vec{X}_p(t')^2\rangle} 
= \exp\left[-\left(\frac{t}{\alpha_p}\right)^{\beta_p}\right],
\end{equation}
where $\alpha_p$ and $\beta_p$ are fitting coefficients.
The longest relaxation time $\tau_1$ is obtained from the time integration of the KWW function of the 1st Rouse mode,
\begin{equation}
 \tau_1=\int_0^{\infty}\exp\left[-\left(\frac{t}{\alpha_1}\right)^{\beta_1}\right] dt =\frac{\alpha_1}{\beta_1}\Gamma\left(\frac{1}{\beta_1}\right),
\end{equation}
where $\Gamma$ represents the gamma function.

\begin{figure}[h]
\centering
\includegraphics[width=.9\hsize]{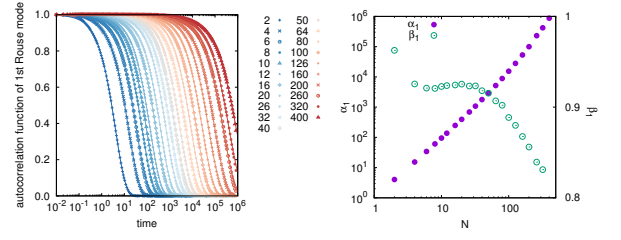}
\caption{Autocorrelation functions of the 1st Rouse mode with the KWW fitting lines (left) and the fitting coefficients of the KWW function $\alpha_1$ and $\beta_1$ (right) between $N=2$ and $N=400$.}
\label{fig.rma}
\end{figure}

\begin{figure}[h]
 \begin{tabular}{c}
\includegraphics[width=.8\hsize]{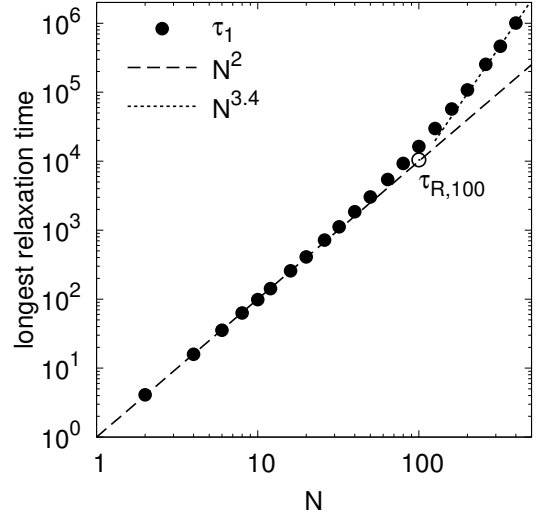}
 \end{tabular}
\caption{The longest relaxation time between $N=2$ and $N=400$.}
\label{fig.tau1}
\end{figure}

Figure~\ref{fig.rma} shows the autocorrelation functions of the 1st Rouse mode with the KWW fitting lines (left) and the fitting coefficients of the KWW function $\alpha_1$ and $\beta_1$ (right) between $N=2$ and $N=400$.
Figure~\ref{fig.tau1} summarizes the longest relaxation time $\tau_1$ between $N=2$ and $N=400$ obtained from the time integration of the KWW function.
The longest relaxation time shows the crossover behavior from the Rouse regime $\tau_1(N)\sim N^2$ to the reptation regime $\tau_1(N)\sim N^{3.4}$.
We have estimated the Rouse relaxation time for $N=100$, $\tau_{\rm R,100}$, by the extrapolation.
The longest relaxation time and the Rouse relaxation time for $N=100$ are determined as $\tau_1=16308$ and $\tau_{\rm R}=10433$.

\section{Automorphism matrices}\label{app.automat}
Here we summarize the automorphism matrix $\tensor{M}$.
According to Dobson~\cite{Dobson2014}, 
\begin{align}
 \tensor{M}_{\rm s}=
\begin{pmatrix}
 1 & 1 & 0 \\
 0 & 1 & 0 \\
 0 & 0 & 1 
\end{pmatrix}
\end{align}
represents the Lees--Edwards boundary conditions (LEBCs)~\cite{LE1972} or the Lagrangian rhomboid boundary conditions (LRBCs)~\cite{Evans1979,HE1994,HT2003} for shear flow and
\begin{align}
 \tensor{M}_{\rm p}=
\begin{pmatrix}
 2 & -1 & 0 \\
 -1 & 1 & 0 \\
 0 & 0 & 1 
\end{pmatrix}
\end{align}
represents the KRBCs~\cite{KR1992,TD1999,BC1999,HT2003} for planar elongational flow.
For uniaxial and biaxial elongational flows, or general elongational flows, the following two automorphism matrices have been found~\cite{Dobson2014}:
\begin{align}
 \tensor{M}_1=
\begin{pmatrix}
 1 & 1 & 1 \\
 1 & 2 & 2 \\
 1 & 2 & 3
\end{pmatrix}
,\quad
 \tensor{M}_2=
\begin{pmatrix}
 2 & -2 & 1 \\
 -2 & 3 & -1 \\
 1 & -1 & 1
\end{pmatrix}
,
\end{align}
where $\tensor{M}_1\cdot\tensor{M}_2=\tensor{M}_2\cdot\tensor{M}_1$. 
Because these matrices are commutative, $\tensor{M}_1$ and $\tensor{M}_2$ are simultaneously diagonalizable.
Solving the eigenvalue equation \eqref{eq.eigen}, we can obtain the reproducible basis vectors $\vec{b}_i^0$ for general elongational flows.
These two matrices $\tensor{M}_1$ and $\tensor{M}_2$ have the same eigenvalues and the same eigenvectors, whereas the correspondence between the eigenvalues and the eigenvectors is different.
The $j$-th eigenvector of $\tensor{M}_1$ corresponds to the $(j+1)$-th eigenvector of $\tensor{M}_2$. (The third eigenvector of $\tensor{M}_1$ corresponds to the first eigenvector of $\tensor{M}_2$.)
The general elongational flows for a long time period have been successfully handled by switching between the two reproducible lattices represented by $\tensor{M}_1$ and $\tensor{M}_2$. 
The other choice for the automorphism matrix has been given by Hunt~\cite{Hunt2015}:
\begin{align}
\tensor{M}_3=
 \begin{pmatrix}
  0 & 1 & 0 \\
  0 & 0 & 1 \\
  1 & -5 & 6
 \end{pmatrix}.
\end{align}
These three matrices, $\tensor{M}_1$, $\tensor{M}_2$, and $\tensor{M}_3$, 
satisfy the following common eigenvalue equation with $k=6$ and $m=5$:
\begin{align}
 p(\lambda)=\lambda^3 - k \lambda^2 + m \lambda -1 =0,
\label{eq.eig.p}
\end{align}
where $\lambda$ is the eigenvalue of the matrix $\tensor{M}_i$.
Each coefficient in Eq. \eqref{eq.eig.p} is related to the matrix $\tensor{M}_i$:
$k={\rm Tr}\tensor{M}_i$, $m=\{({\rm Tr}\tensor{M}_i)^2-{\rm Tr}(\tensor{M}_i^2)\}/2$, and ${\rm det}\tensor{M}_i=1$.
The pairs $(k,m)$ must be integers for the reproducible lattice and lie in the region $m \le k^2/4$ and $k \le m^2/4$ for general extensional flows~\cite{KR1992}.
The polynomial $p(\lambda)$ has a local minimum at 
\begin{equation}
\lambda_0= \frac{k+\sqrt{k^2-3m}}{3}
\end{equation}
when $k^2>3m$.
If the integer pair $(k,m)$ satisfies $p(\lambda_0)=0$,
time-periodic uniaxial and biaxial elongational flows are available.
However, there are no integer pairs $(k,m)$ satisfying $p(\lambda_0)=0$~\cite{KR1992}.
Because the integer pair $(k=6,m=5)$ lies close to the line $k=m^2/4$, the lattice under uniaxial elongational flow is reproducible at the sacrifice of time or strain periodicity.
We can find several matrices with integer components that satisfy Eq. \eqref{eq.eig.p}.
Note that
the initial basis vectors $\vec{b}_i^0$ depend on the automorphism matrix that we have chosen.

\section{Semaev's algorithm}\label{app.semaev}
Semaev's algorithm~\cite{Semaev} has been implemented in the UEF method~\cite{NR2016}. 
This algorithm is more efficient than the conventional lattice reduction algorithm, which is called as the Gaussian reduction algorithm or the LLL-algorithm~\cite{LLL}.
The Gaussian reduction algorithm is based on the Gram-Schmidt projection scheme.
Now, we consider two basis vectors $\vec{b}_1$ and $\vec{b}_2$ where $|\vec{b}_1| \le |\vec{b}_2|$ and reduce them to ${\vec{b}_1}'$ and ${\vec{b}_2}'$ where $|{\vec{b}_1}'| \le |{\vec{b}_2}'|$.
From the Gram-Schmidt process, we can find a new vector $\vec{a}$ as follows:
\begin{align}
\vec{a}&=\vec{b}_2- x_1\vec{b}_1, \\
 x_1 &=\left|\left| \frac{b_{21}}{b_{11}}\right|\right| ,
\end{align}
where $b_{ij}=\vec{b}_i\cdot\vec{b}_j$ and $||\cdot||$ denotes the nearest integer function.
If $|\vec{b}_1| < |\vec{a}| < |\vec{b}_2|$, set $\vec{b}_2=\vec{a}$ and repeat the above process.
Else if $|\vec{a}| \le |\vec{b}_1|$, the reduced vectors are determined as ${\vec{b}_1}'=\vec{a}$ and ${\vec{b}_2}'=\vec{b}_1$. 
When $|\vec{a}|\ge |\vec{b}_2|$, the vectors $\vec{b}_1$ and $\vec{b}_2$ are not reducible, and then we set ${\vec{b}_1}'=\vec{b}_1$ and ${\vec{b}_2}'=\vec{b}_2$.

Since the above algorithm is a pairwise one, 
we have to check all vector pairs $\vec{b}_i$ and $\vec{b}_j$ in three dimension.
Semaev's algorithm~\cite{Semaev} decreases the computational costs.
Now we consider the basis vectors in three dimension $\vec{b}_1, \vec{b}_2$ and $\vec{b}_3$ where $|\vec{b}_1|\le|\vec{b}_2|\le|\vec{b}_3|$.
At first, we reduce the vectors $\vec{b}_1$ and $\vec{b}_2$ using the above Gaussian reduction algorithm.
Then, we reduce $\vec{b}_3$ to $\vec{a}$ if we find a minimum of $|\vec{b}_3+x_2\vec{b}_2+x_1 \vec{b}_1|$ ($\le |\vec{b}_3|$)
with integers $x_1$ and $x_2$.
The reduced vector $\vec{a}$ is represented as
\begin{align}
 \vec{a}=\vec{b}_3+x_2\vec{b}_2+x_1 \vec{b}_1.
\end{align}
The integers $x_1$ and $x_2$ should satisfy $y_1-1 \le x_1 \le y_1 + 1$ and $y_2-1 \le x_2 \le y_2 + 1$, where
\begin{align}
 y_2&=\frac{b_{12}b_{13}-b_{11}b_{23}}{b_{11}b_{22}-b_{12}b_{12}},\\
 y_1&=\frac{b_{12}b_{23}-b_{22}b_{13}}{b_{11}b_{22}-b_{12}b_{12}}.
\end{align}
The denomenator and the numerators of the above equations have the following vector relationships.
\begin{align}
b_{11}b_{22}-b_{12}b_{12}&=|\vec{b}_1 \times \vec{b}_2|^2, \\
b_{12}b_{13}-b_{11}b_{23}&= \vec{b}_1 \cdot ((\vec{b}_1\times \vec{b}_2)\times \vec{b}_3),\\
b_{12}b_{23}-b_{22}b_{13}&= (\vec{b}_3\times(\vec{b}_1\times \vec{b}_2))\cdot \vec{b}_2.
\end{align}
If $|\vec{a}| < |\vec{b}_3|$, then set $\vec{b}_3=\vec{a}$. Replace the subscripts of the basis vectors $\vec{b}_1, \vec{b}_2$, and $\vec{b}_3$ so that $|\vec{b}_1|\le|\vec{b}_2|\le|\vec{b}_3|$. Go back to the Gaussian reduction algorithm for $\vec{b}_1$ and $\vec{b}_2$, and then repeat the above processes until we get $|\vec{a}|\ge |\vec{b}_3|$.
Finally, we will obtain the reduced basis vectors $\vec{b}_1, \vec{b}_2$, and $\vec{b}_3$.

\section{Temperature growth in high-strain rate flow\label{app.temp}}

\begin{figure}[h]
\centering
 \includegraphics[width=.9\hsize]{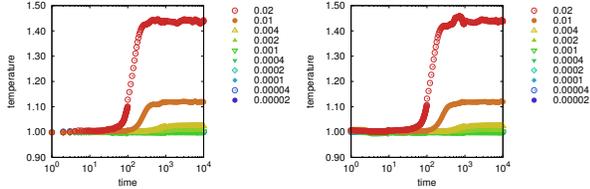} 
 \caption{Time development of temperature in uniaxial elongational flow obtained by the UEF method with $\Delta t = 0.01$ (left) and  $\Delta t = 0.001$ (right). Symbols represent the strain rates $\dot{\epsilon}$ from 0.00002 to 0.02.}
 \label{fig.temp.growth}  
\end{figure}

In this appendix, we discuss the temperature in high strain rates.
Figure~\ref{fig.temp.growth} shows the time development of the temperature in uniaxial elongational flow as discussed in Sec.~\ref{subsec1}.  
When $\dot{\epsilon} \ge 0.002$, the temperature deviates from unity.
The deviation starts from the time when strain hardening appears, as found in Fig.~\ref{fig.uni}, while the temperature remains constant when the strain is small.
The temperature growth apparently correlates with the nonlinear dynamics of polymer chains.
Although we have tested the smaller time width $\Delta t=0.001$ than the reference value $\Delta t=0.01$, we are unable to find any improvements, as shown in the right graph in Fig.~\ref{fig.temp.growth}.

Moreover, we have checked the much smaller time width $\Delta t=0.0001$ and the higher order algorithm for Langevin dynamics~\cite{GJF2013,GJF2014}.
However, these improvements of numerical integration scheme do not prevent the temperature growth.
To understand the cause of the temperature growth, we have investigated molecular states in the steady flows, such as the averaged bond length and the bond orientation tensor, as shown in Appendix~\ref{app.bond}.

\section{Bond length and eigenvalues of bond orientation tensor at steady-state flows}\label{app.bond}

\begin{figure}[h]
 \centering
 \includegraphics[width=.9\hsize]{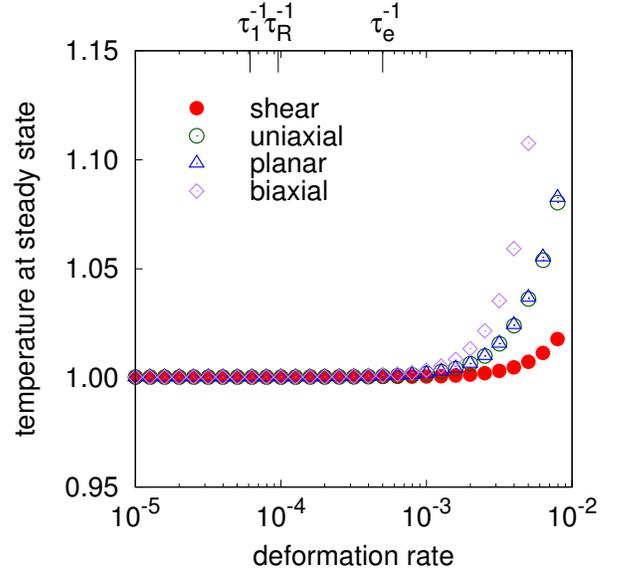} 

\caption{Temperature at steady state in shear flow, uniaxial, planar, and biaxial elongational flows.}
\label{fig.ss.temp}
\end{figure}

In this appendix, we discuss the relationship between the temperature growth in the high-deformation-rate flows and the molecular states.
The temperature at steady-state flows plotted against the deformation rate are summarized in Fig. \ref{fig.ss.temp}.
The temperature growth starts when the deformation rate is higher than $10^{-3}$ which is two times larger than $1/\tau_{\rm e}$. Even in the shear flow, we can observe the temperature growth.

\begin{figure}[h]
 \centering
 \includegraphics[width=.9\hsize]{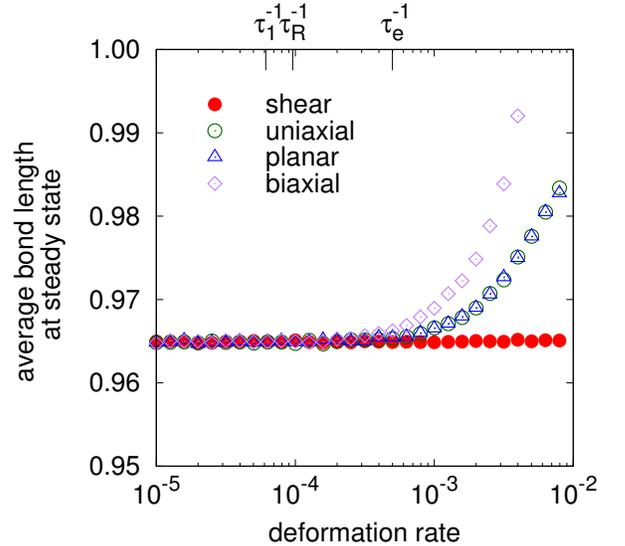} 

\caption{Average bond length at steady state in shear flow, uniaxial, planar, and biaxial elongational flows.}
\label{fig.bond.length}
\end{figure}

At first, we have investigated the average FENE bond length as shown in Fig. \ref{fig.bond.length}.
The average FENE bond length $\langle |\vec{r}| \rangle$ of the Kremer-Grest model in equilibrium is nearly equal to 0.965.
When the deformation rate is higher than $1/\tau_{\rm e}$, the average bond length increases in the elongational flows, while it is kept constant in the shear flow.
From the discrepancy between the temperature growth and the bond length growth,
the bond length growth is not the origin of the temperature growth.

\begin{figure}[h]
\centering
\includegraphics[width=.9\hsize]{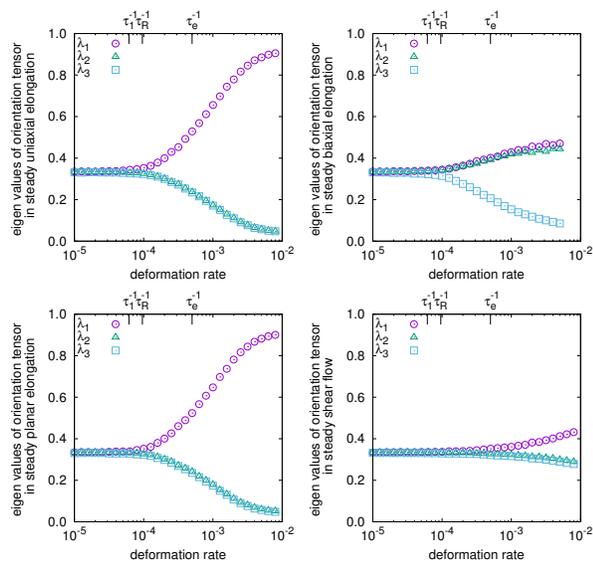} 
\caption{Eigenvalues of bond orientation tensor at steady states in uniaxial elongation (left top),
biaxial elongation (right top), planar elongation (left bottom), and shear flow (right bottom).}
\label{fig.bond.orientation}
\end{figure}

Next, we have investigated the normalized bond orientation tensor $\tensor{O}=\langle \vec{n} \vec{n} \rangle$, where $\vec{n}=\vec{r}/|\vec{r}|$.
The eigenvalues $\lambda_i$ ($i=1, 2, 3$) of the bond orientation tensor $\tensor{O}$ have the following physical meanings:
An isotropic state has $\lambda_1  = \lambda_2 = \lambda_3= \frac{1}{3}$,
a planar isotropic state has $\lambda_1 = \lambda_2 =\frac{1}{2}$ and $\lambda_3=0$, and
an aligned state has $\lambda_1  = 1$ and $\lambda_2=\lambda_3 = 0$.
Because of the excluded volume effect, the ideal aligned state and the ideal planar isotropic state cannot be realized.
The eigenvalues of the normalized bond orientation tensor at steady-state flows are summarized in Fig. \ref{fig.bond.orientation}.
All flows represent the isotropic state when the deformation rate is small.
The uniaxial and planar elongational flows in the high deformation rate exhibit the highly aligned states.
The biaxial elongational flow in the high deformation rate shows the planar isotropic state and the shear flow in the high deformation rate shows the weakly aligned state.
Comparing Fig. \ref{fig.ss.temp} and Fig. \ref{fig.bond.orientation}, 
we can find the inflection point or the maximum slope 
of each eigenvalue appears around the start point of the temperature growth.
When the deformation rate is very high, the FENE bonds are forced to align whereas the excluded volume effect disturbs the bond orientation.
As a result, the LJ particles collides strongly in the high-deformation-rate flows, and therefore the 
excessive supply of energy causes the temperature growth.

%

\end{document}